\documentstyle[12pt,amsfonts]{article}
\parskip=\medskipamount

\def\appendix#1{
  \addtocounter{section}{1}
  \setcounter{equation}{0}
  \renewcommand{\thesection}{\Alph{section}}
 \section*{Appendix \thesection\protect\indent
 \parbox[t]{11.715cm} {#1}}
 \addcontentsline{toc}{section}{Appendix \thesection\ \ \ #1}
  }
\renewcommand{\thefootnote}{\fnsymbol{footnote}}

\newcommand {\cD}{{\cal D}}

\newcommand {\cF}{{\cal F}}

\newcommand {\cH}{{\cal H}}

\newcommand {\cJ}{{\cal J}}

\newcommand {\cM}{{\cal M}}
\newcommand {\cN}{{\cal N}}
\newcommand {\cO}{{\cal O}}

\newcommand {\cV}{{\cal V}}
\newcommand {\cW}{{\cal W}}

\def\a{\alpha}
\def \bi{\bibitem}

\def\b{\beta}

\def\d{\delta}
\def\e{\epsilon}

\def\g{\gamma}
\def\G{\Gamma}

\def\m{\mu}

\def\p{\pi}
\def\q{\theta}
\def\r{\rho}
\def\s{\sigma}
\def\t{\tau}

\def\z{\zeta}
\def\D{\Delta}

\def\J{\Psi}
\def\L{\Lambda}
\def\O{\Omega}

\def\S{\Sigma}
\def\U{\Upsilon}

\newcommand{\ad}{{\dot{\alpha}}}                           
\newcommand{\bd}{{\dot{\beta}}}                            
\newcommand{\ve}{\varepsilon}                            

\newcommand{\pa}{\partial}                           
\newcommand{\hf}{\frac12}

\newcommand{\be}{\begin{equation}}
\newcommand{\ee}{\end{equation}}
\newcommand{\bea}{\begin{eqnarray}}
\newcommand{\eea}{\end{eqnarray}}
\newcommand{\non}{\nonumber}
\begin{document}

\begin{titlepage}
\thispagestyle{empty}

\begin{flushright}
hep-th/0105121 \\
May, 2001
\end{flushright}
\vspace{5mm}

\begin{center}
{\Large \bf Hypermultiplet effective action:\\
\mbox{$\cN = 2$}  superspace approach}
\end{center}
\vspace{3mm}

\begin{center}
{\large
S.M. Kuzenko and I.N. McArthur
}\\
\vspace{2mm}

${}$\footnotesize{
{\it
Department of Physics, The University of Western Australia\\
Crawley, W.A. 6009. Australia}
} \\
{\tt  kuzenko@cyllene.uwa.edu.au},~
{\tt mcarthur@physics.uwa.edu.au}
\vspace{2mm}

\end{center}
\vspace{5mm}

\begin{abstract}
\baselineskip=14pt
In an earlier paper (hep-th/0101127), we developed
heat kernel techniques in $\cN=2$ harmonic superspace
for the calculation of the low-energy effective action
of $\cN=4$ SYM theory, which can be considered as the most
symmetric $\cN=2$ SYM theory. Here, the results are extended
to generic $\cN=2$ SYM theories. This involves a prescription
for computing the variation of the hypermultiplet effective action.
Integrability of this variation allows the hypermultiplet
effective action to be deduced. This prescription permits
a very simple superfield derivation of the perturbative
holomorphic prepotential. Explicit calculations of
the prepotential and the leading non-holomorphic correction
are presented.
\end{abstract}

\vfill
\end{titlepage}

\newpage
\setcounter{page}{1}

\renewcommand{\thefootnote}{\arabic{footnote}}
\setcounter{footnote}{0}

\noindent
{\bf 1.} In conventional quantum field theory,
powerful techniques have been developed for the computation
of low-energy effective actions. These often involve a combination
of (i) the {\it background field method} to allow
the maintenance of manifest gauge invariance throughout
the calculation; and (ii) {\it heat kernel techniques},
which effectively allow in a single step the summation
of an infinite set of Feynman diagrams with increasing
number of insertions of the background field.

Many of the remarkable properties of supersymmetric Yang-Mills
theories, and indeed of superstring theories, are related
to their supersymmetry. In the computation of low-energy
effective actions for these theories, the challenge is therefore
to use the background field method and heat kernel techniques
in a manner which also preserves manifest supersymmetry.
This is most efficiently achieved by formulating the theory
in an appropriate superspace. In the case of $\cN=1$ supersymmetric
Yang-Mills theories, their gauge structure
 can be incorporated into the superspace
formulation in a remarkably simple and elegant geometric manner
(see \cite{WB} for a review). The background field method
and heat kernel techniques for these theories are well-developed
(see \cite{GGRS,BK} for reviews).

{}For $\cN=2$ supersymmetric Yang-Mills theories,
the harmonic superspace approach \cite{GIKOS,GIOS}
provides a universal setting for the description of their dynamics.
The background field method in harmonic superspace
was elaborated in refs. \cite{BBKO,BK2}.
{}Following this, two important applications of the method were given:
(i) the first rigorous proof of the $\cN=2$
non-renormalization theorem \cite{BKO};
(ii) the harmonic-superspace computation
of the leading non-holomorphic quantum corrections
in $\cN=4$ super Yang-Mills theory \cite{BK2,BBK}.
However, until recently, heat kernel techniques in $\cN=2$ superspace
remained almost totally undeveloped, with the result
that very little has been achieved in the way of explicit
computations of low-energy effective actions for
$\cN=2$ supersymmetric Yang-Mills theories.
A significant development in this direction was achieved in \cite{KM}.
{}For the one-loop effective action
of  the $\cN=4$ super Yang-Mills theory
(which can be viewed the most symmetric $\cN=2$ model),
we obtained an $\cN = 2$ superfield
representation which is free of coinciding harmonic
singularities and which permits
a straightforward evaluation of low-energy
quantum corrections in the framework of an $\cN=2$
superfield heat kernel technique.

In the present letter, we extend the one-loop
results of \cite{KM} to the case of generic $\cN=2$ super
Yang-Mills theories. Essentially, such an extension is equivalent
to developing heat kernel techniques for
computing the effective action
of a hypermultiplet coupled to a background $\cN=2$ vector
multiplet. Before proceeding,
we justify this statement and sketch
the most important aspects of the harmonic superspace formulation
in this context.

The $\cN=2$ harmonic superspace ${\Bbb R}^{4|8}\times S^2$
extends conventional superspace,
with coordinates $z^M = (x^m , \q^\a_i , {\bar \q}_\ad^i )$,
 by the two-sphere $S^2 =SU(2)/U(1)$
parametrized by harmonics, i.e., group
elements
\bea
({u_i}^-\,,\,{u_i}^+) \in SU(2)~, \quad
u^+_i = \ve_{ij}u^{+j}~, \quad \overline{u^{+i}} = u^-_i~,
\quad u^{+i}u_i^- = 1 ~.
\eea
The main conceptual advantage of harmonic superspace
is that the $\cN=2$ vector multiplet and
hypermultiplets can be described by {\it unconstrained} superfields
over the analytic
subspace of ${\Bbb R}^{4|8}\times S^2$
 parametrized by the variables
$ \z^\cM \equiv (x^m_A,\q^{+\a},{\bar\q}^+_{\dot\a}, \,
u^+_i,u^-_j) $,
where the so-called analytic basis is defined by
\be
x^m_A = x^m - 2{\rm i} \q^{(i}\s^m {\bar \q}^{j)}u^+_i u^-_j~, \qquad
 \q^\pm_\a=u^\pm_i \q^i_\a~, \qquad {\bar \q}^\pm_{\dot\a}=u^\pm_i{\bar
\q}^i_{\dot\a}
\ee
and represents a generalization of the chiral superspace basis in
$\cN=1$ supersymmetry. The $\cN=2$ vector multiplet is described
by a real analytic superfield $\cV^{++} = \cV^{++}_I (\z) \,T_I$
taking its values in the Lie algebra of the gauge group.
A hypermultiplet transforming
in a representation $R$ of the gauge group
is described by an analytic superfield
$q^+ (\z)$ and its conjugate $\breve{q}^+ (\z)$.
The classical action
for a generic $\cN=2$ super Yang-Mills theory is
\be
S_{\cN=2 \,{\rm YM} }~ =~
\frac{1 }{2 g^2 }
 \int {\rm d}^4 x
{\rm d}^4 \q \; {\rm tr}\,  \cW^2
 - \int  {\rm d} \zeta^{(-4)}\,
\breve{q}{}^+ \cD^{++}q^+ ~,
\label{action}
\ee
where $\cW$ is the $\cN=2$ covariantly
chiral superfield strength \cite{GSW},
$ {\rm d} \zeta^{(-4)}$ denotes the analytic subspace
integration measure, and $\cD^{++} = D^{++} +{\rm i} \, \cV^{++}$
is the analyticity-preserving covariant derivative.
The first term in (\ref{action}), which is
the action of the $\cN=2$ pure super Yang-Mills
theory \cite{GSW}, can be expressed as a gauge-invariant functional
of $\cV^{++}$ \cite{Z}.
The second term in (\ref{action}) is the action
of a massless hypermultiplet. The massive case simply corresponds
to allowing a constant expectation value for $\cV^{++}$ along
an Abelian subalgebra of the gauge algebra
(see, e.g., \cite{BBIKO,DK,IKZ}).

The hypermultiplet effective action $\G_{\rm H}^{(R)}$ is defined by
\be
\exp \Big( {\rm i} \, \G_{\rm H}^{(R)} [\cV^{++}] \Big)
= \int [{\rm d} \breve{q}^+] \, [{\rm d} q^+]\,
\exp \Big(-{\rm i} \int  {\rm d} \zeta^{(-4)}\,
\breve{q}{}^+ \cD^{++}q^+  \Big)~.
\label{path}
\ee
It can be shown \cite{BK2} that the one-loop effective action
of the theory (\ref{action}) is
\be
\G_{\cN=2 \,{\rm YM} }~ =~ \G_{\cN=4 \,{\rm YM} }
~+~ \G_{\rm H}^{(R)} ~-~ \G_{\rm H}^{(ad)}~,
\ee
with $\G_{\cN=4 \,{\rm YM} }$ the one-loop effective action
of $\cN=4$ super Yang-Mills theory. The low-energy
structure of $\G_{\cN=4 \,{\rm YM} }$ has been studied in
\cite{KM}. Therefore, it remains to analyse the hypermultiplet
effective action. It is worth pointing out that here
we are only interested in the functional dependence of the
effective action on the $\cN=2$ vector multiplet, and ignore
all quantum corrections with hypermultiplet external legs.
The latter were discussed, to some extent, in
\cite{IKZ,GR}.

The outline of the paper is as follows.
In section 2, an expression for the variation of the
hypermultiplet effective action is derived, and this is
used in section 3 to calculate the perturbative holomorphic
prepotential.  The technique for computing non-holomorphic
higher-derivative quantum corrections is described in section 4.
This is followed by a short conclusion.

\noindent
{\bf 2.} As determined by (\ref{path}), the
formal definition of the hypermultiplet effective action
reads
\be
\G_{\rm H} = {\rm i} \,{\rm Tr}\, \ln \cD^{++}
= - {\rm i} \,{\rm Tr}\, \ln G^{(1,1)}~,
\label{formal}
\ee
with $G^{(1,1)}(\z_1 , \z_2) $  the hypermultiplet Green function:
\bea
\cD^{++}_1 G^{(1,1)}(\z_1 , \z_2)   &=& \d_A^{(3,1)}(\z_1 , \z_2)
~, \non \\
G^{(1,1)}(\z_1 , \z_2) &=& - \frac{1}{{\stackrel{\frown}{\Box}}{}_1}
(\cD_1^+)^4 \,(\cD_2^+)^4
 \delta^{12}(z_1-z_2)
{1\over (u^+_1 u^+_2)^3}~,
\eea
where ${}{\stackrel{\frown} {\Box}}{}$ is the analytic d'Alembertian
\bea
{\stackrel{\frown}{\Box}}{}&=&
{\cal D}^m{\cal D}_m+
\frac{{\rm i}}{2}({\cal D}^{+\a}\cW){\cal D}^-_\a+\frac{{\rm i}}{2}
({\bar{\cal D}}^+_{\dot\alpha}{\bar \cW}){\bar{\cal D}}^{-{\dot\alpha}}-
\frac{{\rm i}}{4}({\cal D}^{+\a} {\cal D}^+_\a \cW) \cD^{--}\non \\
&{}& +\frac{{\rm i}}{8}[{\cal D}^{+\alpha},{\cal D}^-_\alpha] \cW
+ \frac{1}{2}\{{\bar \cW},\cW \}~.
\eea
The above definition is purely formal, since the operator
$\cD^{++}$ in (\ref{formal}) maps the space of analytic
superfields $q^+$ with $U(1)$ charge $+1$ onto a space of analytic
superfields possessing $U(1)$ charge $+3$, and therefore
its determinant ${\rm Det} \,\cD^{++}$ is ill-defined\footnote{This
is similar to the well-known situation for chiral fermions.}.
However, the expression for an arbitrary variation
of the effective action
\be
\d \G_{\rm H} = - {\rm Tr}\,\Big\{\d \cV^{++} \,G^{(1,1)}\Big\}
\label{var1}
\ee
is well-defined.
On formal grounds, this variation is integrable, since
\be
\d_2 \, \d_1  \G_{\rm H} = {\rm i}\,
{\rm Tr}\,\Big\{\d \cV_1^{++} \,G^{(1,1)}\,
\d \cV_2^{++} \,G^{(1,1)}  \Big\} = \d_1 \, \d_2  \G_{\rm H} ~.
\ee
Our goal in the present paper is first to develop
a covariant heat kernel technique
for computing $\d \G_{\rm H}$, and then to integrate this
variation to yield $\G_{\rm H}$.

Using Schwinger's proper-time
representation,
\be
\frac{1}{ {\stackrel{\frown}{\Box}}}= {\rm i}
\int_{0}^{\infty}{\rm d}s\;
{\rm e}^{-{\rm i}\,s\,{\stackrel{\frown}{\Box}}}~,
\ee
we introduce a regularized variation of the effective action
\bea
\d\G_{{\rm H}, \ve} &=&
\m^{2 \ve}\, {\rm tr} \int_{0}^{\infty}{\rm d}({\rm i}s)\,
({\rm i}s)^\ve  \int {\rm d} \z^{(-4)}\, \d \cV^{++}
\non \\
&& \qquad \qquad \times \,
{\rm e}^{-{\rm i}\,s\,{\stackrel{\frown}{\Box}}_1} \,
(\cD_1^+)^4 \,(\cD_2^+)^4 \,
\frac{ \delta^{12}(z_1-z_2)}{ (u^+_1 u^+_2)^3 }
\Big|_{1 = 2}~,
\eea
with $\ve$ the ultraviolet regularization parameter,
set to zero at the end of calculations,
and $\m$ the normalization point.
The expression for $\d \G_{{\rm H}, \ve}$ can be brought to a more
useful form by applying the identity \cite{KM}
\bea
&& (\cD^+_1)^4 (\cD^+_2)^4 \;
\frac{\d^{12} (z_1 -z_2)  }{(u^+_1 u^+_2)^3}
\label{master}   \\
&& \qquad =
(\cD^+_1)^4 \;
\left\{ (\cD^-_1)^4 \,(u^+_1 u^+_2)
- \frac{\rm i}{2} \,
\D^{--}_1\;
(u^-_1 u^+_2)
- {\stackrel{\frown}{\Box}}_1 \;
\frac{(u^-_1 u^+_2)^2 }{(u^+_1 u^+_2)} \right\}
\;\d^{12} (z_1 -z_2)  ~,
\non
\eea
where
\bea
\D^{--} =\cD^{\a \ad} \cD^-_{\a} {\bar \cD}^-_{\ad}
&+& \hf \cW (\cD^-)^2 + \hf {\bar \cW} ({\bar \cD}^-)^2 \non \\
&+& (\cD^- \cW) \cD^- + ({\bar \cD}^- {\bar \cW}) {\bar \cD}^-
+\hf (\cD^- \cD^- \cW) ~.
\eea

The two-point function in the first line of (\ref{master})
contains a harmonic distribution\footnote{See ref. \cite{GIOS}
for a detailed discussion of harmonic delta-functions and
harmonic distributions of the general form $(u^+_1 u^+_2)^{-n}$,
where $n>0$.} which is singular at coincident points, $u_1 = u_2$.
On the right hand side of (\ref{master}), it is only the third term
which contains a potential coinciding harmonic singularity.
However, this singular term does not
contribute to $\d \G_{{\rm H}, \ve}$, and therefore
$\d \G_{{\rm H}, \ve}$ is free of harmonic singularities.
To see this, note  that the expression
\be
\U_\ve = \m^{2 \ve}\, {\rm tr} \int_{0}^{\infty}{\rm d}({\rm i}s)\,
({\rm i}s)^\ve \,
{\rm e}^{-{\rm i}\,s\,{\stackrel{\frown}{\Box}}_1} \,
(\cD_1^+)^4 \, {\stackrel{\frown}{\Box}}_1 \,
 \delta^{12}(z_1-z_2)\Big|_{z_1 = z_2}~,
\ee
contains no divergences in $\ve$.
This fact and the identity $(\cD^+)^4 \, {\stackrel{\frown}{\Box}}
= {\stackrel{\frown}{\Box}} \, (\cD^+)^4 $ then imply
\bea
\lim_{\ve \to 0}\, \U_\ve &=&  {\rm tr}
\int_{0}^{\infty}{\rm d}({\rm i}s)\,
{\rm e}^{-{\rm i}\,s\,{\stackrel{\frown}{\Box}}_1} \,
{\stackrel{\frown}{\Box}}_1 \,(\cD_1^+)^4 \,
 \delta^{12}(z_1-z_2)\Big|_{z_1 = z_2} \non \\
&=& {\rm tr} \;  (\cD_1^+)^4 \,
 \delta^{12}(z_1-z_2)\Big|_{z_1 = z_2} =0~,
\eea
since at least eight spinor derivatives are required
to produce a non-vanishing result when acting on
the Grassmann delta-function $\d^8(\q_1 -\q_2)$
before  setting $\q_1 =\q_2$. As a result,
we arrive at the following expression for $\d \G_{{\rm H}, \ve}$:
\bea
\d \G_{{\rm H}, \ve}
& \equiv &  {\rm tr} \int {\rm d} \z^{(-4)}\, \d
\cV^{++} \cJ_\ve^{++}
= \m^{2 \ve}\, {\rm tr} \int_{0}^{\infty}{\rm d}({\rm i}s)\,
({\rm i}s)^\ve  \int {\rm d} \z^{(-4)}\, \d \cV^{++}
{\rm e}^{-{\rm i}\,s\,{\stackrel{\frown}{\Box}}_1} \non \\
& \times & (\cD_1^+)^4 \,
\left\{ (\cD^-_1)^4 \,(u^+_1 u^+_2)
- \frac{\rm i}{2} \,
\D^{--}_1\;
(u^-_1 u^+_2) \right\}\, \delta^{12}(z_1-z_2)
\Big|_{1 =2} ~.
\label{master2}
\eea
This is our working formula for the computation of
$\d \G_{{\rm H}, \ve}$.

The effective current $\cJ_\ve^{++}$ in (\ref{master2})
should be analytic,
\be
\cD^+_\a \cJ_\ve^{++} = {\bar \cD}^+_\ad \cJ_\ve^{++} =0~.
\ee
The requirement of gauge invariance of the effective action
is equivalent to the conservation equation
\be
\cD^{++} \, \cJ_\ve^{++} =0~.
\ee

Relation (\ref{master2}) simplifies considerably when
the background gauge superfield satisfies the classical
equation of motion
\be
\cD^{(i} \cD^{j)}\, \cW
= {\bar \cD}^{(i} {\bar \cD}^{j)} \,{\bar \cW} = 0~.
\label{on-shell}
\ee
Then $\d \G_{{\rm H}}$ turns out to be free of
ultraviolet divergences, and eq. (\ref{master2}) takes the form
\bea
\d \G_{{\rm H}}
= -\hf \, {\rm tr} \int_{0}^{\infty}{\rm d}s
  \int {\rm d} \z^{(-4)}\, \d \cV^{++}
{\rm e}^{-{\rm i}\,s\,{\stackrel{\frown}{\Box}}_1} \,
\D^{--}_1\, (\cD_1^+)^4 \,
\delta^{12}(z_1-z_2)
\Big|_{z_1 = z_2} ~.
\label{master3}
\eea
This simplified variation is useful for
the computation of non-holomorphic corrections
to the effective action, which will be done
in section 4.

\noindent
{\bf 3.} To this point, our results are
applicable for an arbitrary
non-Abelian gauge group. In the remainder of the paper,
we specialize to the case of a $U(1)$ gauge group in order
to illustrate the techniques for computing the effective
action. The Abelian gauge field and chiral field strength
will be denoted by $V^{++}$ and $W$ respectively.

In the Abelian case, quantum corrections to the effective action
can be computed in the framework of a derivative expansion.
On general grounds, the part of $\G_{\rm H}$ which does
not contain space-time derivatives of $W$ and $\bar W$
should have the following structure \cite{BKT}:
\bea
\G_{\rm H} &=& \int {\rm d}^4 x
{\rm d}^4 \q \, \cF (W ) ~+~
\int {\rm d}^4 x
{\rm d}^4 {\bar \q} \, {\bar \cF} (\bar W )  \label{structure}\\
&+& \int {\rm d}^4 x  {\rm d}^4 \q {\rm d}^4 {\bar \q} \,
\Big\{ c \,\ln \frac{W}{\L} \, \ln \frac{\bar W}{\L}
~+~ \ln \frac{W}{\L} \, {\bar \S} (\bar \J ) +
\ln \frac{\bar W}{\L} \, \S ( \J )
~+~ \O (\J, \bar \J ) \Big\} ~, \non
\eea
where
$$
{\bar { \J}} = \frac{1}{ {\bar W}^2}\,
D^4 \ln \frac{W}{\L}~, \qquad
{\J} = \frac{1}{  W^2}\,
{\bar D}^4 \ln \frac{\bar W}{\L}~,
$$
with $\L$ a formal scale which in fact drops out
from all structures listed.
Here $\cF$ is a holomorphic function
(often called the perturbative prepotential)
which is the starting point of the Seiberg-Witten theory \cite{SW};
it encodes the quantum corrections with at most two derivatives
in components (super $F^2$ terms).
The first term in the second line of (\ref{structure})
comprises the leading higher-derivative quantum corrections
(super $F^4$ terms)
and  is known to be one-loop exact \cite{DS}. Finally,
the holomorphic ($\S$) and real analytic ($\O$) functions
determine higher-derivative quantum corrections
(super $F^6$ and higher order terms).
We will demonstrate how
these quantum corrections can be computed in the framework
of equation (\ref{master2}).

It is worth making a few comments on the explicit structure
of the low-energy action (\ref{structure}).
The classical action (\ref{action}) is invariant under
the $\cN=2$ superconformal group (see \cite{superconformal}
for a discussion of superconformal transformations
in harmonic superspace). The superconformal symmetry is
known to be anomalous at the quantum level, and the anomaly
is completely determined by the holomorphic prepotential.
The non-holomorphic part of the effective action should be
superconformally invariant. This dictates the structure
in the second line of (\ref{structure}); see \cite {BKT}
for more details.

We first turn to the computation of holomorphic quantum corrections
to $\G_{\rm H}$ of the general form
$S_{\rm eff} =  {\rm Re}\,\int {\rm d}^4 x
{\rm d}^4 \q \, \cF (W ).$
Under an arbitrary variation
$V^{++} \to V^{++} + \d V^{++}$ of the analytic gauge field,
\be
\d S_{\rm eff} = {1\over 4}
\int {\rm d} \z^{(-4)}\, \d V^{++}
\Big\{ (D^+)^2 \, \cF'(W) ~+~
( {\bar D}^+)^2 \,{\bar \cF}' ({\bar W}) \Big\}~,
\label{hol2}
\ee
see, e.g., \cite{DK} for a derivation.
Eq. (\ref{hol2}) indicates that to compute
$\cF(W)$, one has to retain all terms in  $\d \G_{{\rm H}, \ve}$
which involve exactly two spinor derivatives
of the field strengths.

Both terms in (\ref{master2}) contribute nontrivially
to $\d S_{\rm eff}$. The first term contains exactly
eight spinor derivatives to annihilate
the Grassmann delta-function, via the identity
\be
(D^+)^4\,(D^-)^4\, \d^8(\q-\q')\Big|_{\q=\q'}=1~.
\ee
Due to harmonic identities $(u^+_1 u^+_2)|_{1=2} =0$
and $(u^-_1 u^+_2)|_{1=2} =-1$, this term produces
a non-vanishing contribution only if we pick up a factor
of  $(D^+ D^+ W)\, \cD^{--}$ in the decomposition of
$\exp (-{\rm i}\,s\,{\stackrel{\frown}{\Box}}_1)$
and use it to act on $(u^+_1 u^+_2)$ (the harmonic
derivative $D^{--}$ is defined by $D^{--} u^+ = u^-$,
$ D^{--} u^- = 0$).
Since we are working to second order in spinor derivatives,
it is sufficient for our purporses to approximate
$\exp (-{\rm i}\,s\,{\stackrel{\frown}{\Box}})
\approx 1/4 \, (D^+ D^+ W)\,
\exp (-{\rm i}\,s\, W \bar W ) \,
\exp (-{\rm i}\,s\,\pa^a \pa_a) \,D^{--}$.
As a result, the first term in (\ref{master2})
contributes as follows (with the analytic subspace integral
omitted):
\bea
&& \frac{ \m^{2\ve} }{ (8\p)^2 }\,
\d V^{++} \,(D^+ D^+ W)
\int_{0}^{\infty}{\rm d}\t \t^{\ve -1}\,
{\rm e}^{-\t W \bar W }
\label{con1} \\
&=& \frac{1 }{ (8\p)^2 }\,
\d V^{++} \Big( \frac{1}{\ve} \,(D^+ D^+ W)
-\ln \frac{W}{\m} (D^+ D^+ W)
- \ln \frac{\bar W}{\m} ({\bar D}^+ {\bar D}^+ {\bar W})
\Big)
 ~+~\cO ( \ve )~, \non
\eea
where we have applied Schwinger's rotation
$s \rightarrow -{\rm i}\, \t$ and used the Bianchi
identity \cite{GSW}
\be
D^+ D^+ \, W = {\bar D}^+ {\bar D}^+ \bar W ~.
\ee
It is worth pointing out that the $1/\ve$ term in
(\ref{con1}) completely determines
the divergent part of $\G_{\rm H}$; there are no other contributions.

In order for the second term in (\ref{master2})
to contribute to $\d  S_{\rm eff}$, we have to
accumulate a product of two spinor derivatives
from the expansion of
$\exp (-{\rm i}\,s\,{\stackrel{\frown}{\Box}})$
to have enough spinor derivatives to
annihilate the Grassmann delta-function. The result reads
\bea
 -\frac{1 }{ (8\p)^2 }\,
\d V^{++} \Big(
{\bar W}\, D^+ W D^+ W
&+& W\,{\bar D}^+ {\bar W} {\bar D}^+ {\bar W} \Big)
\int_{0}^{\infty}{\rm d}\t \,
{\rm e}^{-\t W \bar W }
~+~\cO ( \ve )   \non \\
= -\frac{1 }{ (8\p)^2 }\,
\d V^{++}
\Big(
\frac{ D^+ W D^+ W}{W} &+&
\frac{ {\bar D}^+ {\bar W} {\bar D}^+ {\bar W} }
{ \bar  W }
\Big)
~+~\cO ( \ve )~.
\label{con2}
\eea

{}From eqs. (\ref{hol2}), (\ref{con1}) and (\ref{con2})
we read off the divergent part of the effective action
\be
\G_{{\rm H}, \, {\rm div}} = \frac{1 }{ 32\p^2 \,\ve}
 \int {\rm d}^4 x
{\rm d}^4 \q \, W^2
\ee
and the perturbative holomorphic prepotential
\be
\cF ( W ) =  - \frac{1 }{ 32\p^2 }\, W^2
 \ln \frac{W }{\m}~.
\label{holprep}
\ee
The low-energy holomorphic action of the charged
hypermultiplet, which is generated by $\cF ( W )$,
was previously computed in \cite{BBIKO} by using
the standard $\cN=2$ harmonic superspace Feynman rules \cite{GIOS},
and this result was then extended in \cite{hol} to the case of generic
$\cN=2$ super Yang-Mills models on the Coulomb branch.
In the present paper, we have provided the first derivation
of $\cF ( W )$ on the basis of the $\cN=2$ background field
formulation. In the (background non-covariant) approach of
refs. \cite{BBIKO,hol}, the calculation of $\cF ( W )$
is quite involved. As is seen from the above discussion,
the derivation of $\cF ( W )$ in our covariant approach
is almost trivial and, in fact, simpler than
the known $\cN=1$ superfield calculations of effective K\"ahler potentials
\cite{BKY,BK,DGR,PW,PB}.

\noindent
{\bf 4.} To compute the higher-derivative quantum corrections,
which appear in the second line of (\ref{structure}),
it is sufficient to evaluate $\d \G_{\rm H} $ for an
on-shell background vector multiplet, defined by eq. (\ref{on-shell}).
In this case, $\d \G_{\rm H} $ takes the simplified form
(\ref{master3}), and the problem reduces
to computing the kernel\footnote{When restricting the background
vector multiplet to satisfy eq. (\ref{on-shell}),
we completely lose the corrections to $\d \G_{\rm H} $
containing factors of
$D^\pm D^\pm W = {\bar D}^\pm {\bar D}^\pm \bar W$.
The role of such corrections,
however, is just to complete the surviving contributions
in such a way that the effective current $\cJ^{++}$
is analytic.}
(strictly speaking the coincidence limit of a kernel)
\be
K^{--}(z;s) = \lim_{z
\rightarrow z'} {\rm e}^{- {\rm i}\, s\,
{\stackrel{\frown}{\Box}} } \, \D^{--}
(\cD^+)^4 \d^{12}(z-z')~.
\label{kernel}
\ee
Since we ignore the corrections to $\G_{\rm H} $
containing space-time derivatives of $W$ and $\bar W$,
we can further restrict the background vector multiplet
by the conditions
\be
W|_{\q =0} = {\rm const}~, \qquad
D^i_\a W|_{\q =0} = {\rm const}~, \qquad
D^i_{(\a} D_{\b) i} W|_{\q =0} \equiv 8 F_{\a \b}
 = {\rm const}~.
\ee
Then, the kernel (\ref{kernel})
can be computed exactly by adapting techniques developed in
\cite{MG}. Below we outline the main steps in the calculation of
$K^{--}(z;s)$.

Replacing the delta-function by its  Fourier
representation
$$
(\cD^+)^4 \d^{12}(z-z') =  \int {\rm d} \eta^{(+4)} \,
{\rm e}^{{\rm i} \,k_a (x - x')^a}
{\rm e}^{{\rm i}\, \e^- (\q - \q')^+} \,
{\rm e}^{{\rm i}\, \bar \e^{\, -} ( \bar \q - \bar \q')^+}
$$
with
$ \int {\rm d} \eta^{(+4)} = 16 \int \frac{d^4k}{(2 \p)^4} \, \int
{\rm d}^2 \e^- \, \int {\rm d}^2 \bar \e^{\, -}$,
the kernel can be expressed
\bea
K^{--}(z;s) & = & \int {\rm d} \eta^{(+4)} \,
{\rm e}^{- {\rm i}\, s \,
{\stackrel{\sim}{\Box}}
}\,
\Big( X^{ \ad \a} X_{\a}^- \bar X_{\ad}^- +  \frac12 W
(X^-)^2 + \frac12 \bar W ( \bar X^-)^2
\non \\
& + &
(D^{- \a } W) X^-_{\a} +
( \bar D_{\ad}^{ -} \bar W) \bar X^{-\ad }\Big)
\label{kernel2}
\eea
where
\bea
X_a &=& \cD_a + {\rm i} k_a, \,\,\,\,\, X^-_{\a} = \cD^-_{\a} + {\rm i}
\e^-_{\a}, \,\,\,\,\, {\bar X}^-_{\ad} = {\bar \cD}^-_{\ad} + {\rm i}
{\bar \e}^{\,-}_{\ad} ~;\non \\
{\stackrel{\sim}{\Box}} &=& X^a X_a
+ \frac{\rm i}{2} (D^{+\a } W) X^-_{\a}
+\frac{\rm i}{2}
( \bar D_{\ad}^{ +} \bar W) \bar X^{-\ad}
+ W \bar W~. \non
\eea
The ``shifted'' operators $X$'s satisfy the same algebra as the
corresponding operators $\cD$'s.

Commuting $W,$ $\bar W$ and their spinor derivatives
in (\ref{kernel2}) to the left through $\exp (-{\rm i}\, s \,
{\stackrel{\sim}{\Box}}) $,
and noting that $D_{\a}^- W$ and ${\bar D}_{\ad}^- \bar W$
commute with $ {\stackrel{\sim}{\Box}}$  on-shell, we obtain
\bea
K^{--}(z;s) &=&  K_{X^{\ad \a}X_{\a}^-  \bar X_{\ad}^-} (z;s) +
(D^{-\a } W) \,
K_{X_{\a}^-}(z;s) - ({\bar D}^{-\ad } \bar W ) \,
K_{\bar X_{\ad}^-}(z;s) \non \\
& +&  \hf \Big\{ W + \hf \,
(D^+_\a W)\,
\Big( \frac{{\rm e}^{-sN} - {\bf 1}}{N} \Big)^{\a}{}_{\b}\,
(D^{-\b} W)
\Big\} \, K_{(X^-)^2}(z;s) \non
\\ & +&  \hf \Big\{ \bar W - \hf \,
( {\bar D}^+_\ad \bar W )\,
\Big( \frac{{\rm e}^{-s \bar N } - {\bf 1}}{ \bar N} \Big)
^{\ad}{}_{\bd} \, ( \bar D^{- \bd }\bar W)
\Big\} \,
K_{(\bar X^-)^2}(z;s)~,
\label{K1}
\eea
where we have introduced generalized Gaussian moments
$$K_{\hat{O}} (z;s) ~=~ \int {\rm d} \eta^{(+4)} \,
{\rm e}^{- {\rm i} \, s\,
{\stackrel{\sim}{\Box}}
} \, \hat{O}$$
and defined
$$
N^{\a}{}_{\b} ~=~ \hf \,D^{-\a} D_{\b}^+ W~, \qquad \bar N^{\ad}{}_{\bd}
~ =~- \hf \,\bar D^{-\ad} \bar D_{\bd}^+ \bar W~.
$$
Use of  the identity
$ 0 = \int {\rm d} \eta^{(+4)} \, \pa / \pa \,\e_{\a}^- \,
[\, \exp ( - {\rm i}\, s \,
{\stackrel{\sim}{\Box}} ) \,X_{\b}^-  X_{\g}^- \,] $ yields
\be
 K_{X_{ \a }^-} (z;s) = \frac{1}{4} \, (D_{\b}^+ W) \,
\Big( \frac{{\rm e}^{-sN} - {\bf 1}}{N}\Big)^{\b}{}_{\a} \,
K_{(X-)^2}(z;s)~,
\label{Ka}
\ee
and then (\ref{K1}) collapses  to
\be
K^{--}(z;s) =  K_{X^{\ad \a}X_{\a}^- \bar X_{\ad}^-} (z;s) + \frac12 W \,
 K_{(X^-)^2}(z;s)
+  \frac12 \bar W \,  K_{(\bar X^-)^2}(z;s)~.
\label{K2}
\ee
As in the derivation of (\ref{Ka}),
more general identities of the form
\be
0 =  \int {\rm d} \eta^{(+4)} \,\frac{\pa }{ \pa k_{\a \ad}
}
\Big[ {\rm e}^{- {\rm i} \, s\,
{\stackrel{\sim}{\Box}}
} \, \hat{O} \Big] ~, \qquad
0 = \int {\rm d} \eta^{(+4)} \,\frac{\pa }{\pa \e^-_\a
}
\Big[ {\rm e}^{- {\rm i} \, s\,
{\stackrel{\sim}{\Box}}
} \, \hat{O} \Big]
\label{id}
\ee
allow one to express all the moments in (\ref{K2})
via a single one, $K_{(X^-)^2 (\bar X^-)^2}(z;s)$, as follows:
\bea
K_{(X^-)^2}
&=& \frac{1}{16} \,
({\bar D}^+ \bar{W}\,  {\bar D}^{+ } \bar{W})
\,{\rm tr} \Big( \frac{{\bf 1} -  \cosh s \bar{N}}
{\bar{N}^2 } \Big)
\, K_{(X^-)^2
(\bar{X}^-)^2} ~,
\label{K4} \\
 K_{X^{\ad \a}X_{\a}^- \bar X_{\ad}^-}
&=& \frac{1}{16}
(D^+_\a W)\,
\Big( \frac{{\rm e}^{-sN} - {\bf 1}}{N} \Big)^{\a}{}_{\b}\,
( {\bar D}^+_\ad \bar W )\,
\Big( \frac{{\rm e}^{-s \bar N } - {\bf 1}}{ \bar N} \Big)
^{\ad}{}_{\bd} \,
S^{\bd \b} \,  K_{(X^-)^2 (\bar X^-)^2}~,
\non
\eea
where (with
${\rm i} \,F_{\a \ad, \b \bd} = [ \cD_{\a \ad}, \cD_{\b \bd}]$)
\bea
S_{\a \ad}
& = & -\hf \,\sum_{n=1}^{\infty}
\sum_{p=1}^n \, \frac{s^{n+1}}{(n+1)!} \, \Big[ \Big(
\frac{{\rm e}^{s F} -{\bf 1}}{F} \Big)^{-1}
(-F)^{n-p}\Big]_{\a \ad}{}^{\bd \b}
\label{K3}
\\
& \times & \Big\{
(D_{\r}^+ W) \,
( {\rm e}^{-s N}\, N^{p-1} )^{\r}{}_{\b} \, ({\bar D}_{\bd}^- \bar{W})
- \,  ({\bar D}_{\dot{\r}}^+ \bar{W} ) \,
( {\rm  e}^{-s \bar{N}}\,  \bar{N}^{p-1} )^{\dot{\r}}{}_{\bd}
\, (D_{\b}^- W)  \Big\}~.
\non
\eea

The remaining step is the computation of
$K_{(X^-)^2 (\bar{X}^-)^2}(z;s)$.
This proceeds via the differential equation
\bea
{\rm i}\, \frac{\rm d}{{\rm d} s} \, K_{(X^-)^2 (\bar{X}^-)^2}
&=&  \frac{1}{2}
\, K_{X^{\ad \a} \, X_{\a \ad}  \, (X^-)^2 \,(\bar{X}^-)^2 }
+ \int {\rm d} \eta^{(+4)} \,
{\rm e}^{- {\rm i}\, s\,{\stackrel{\sim}{\Box}} }
\, W
\bar{W}  \, (X^-)^2 \,(\bar{X}^-)^2~.
\label{K5}
\eea
The trick is to express the right hand side in terms of
$K_{(X^-)^2 (\bar{X}^-)^2},$ thus  establishing a linear differential
equation for $K_{(X^-)^2 (\bar{X}^-)^2}$.
This is easily done for the second term in (\ref{K5}) by commuting
$W  \bar{W}$ to the left. For the first term, it is necessary
to proceed via the first identity (\ref{id}) with
$\hat{O} = X_{\b \bd}   (X^-)^2 (\bar{X}^-)^2$
and $\hat{O} = (X^-)^2 (\bar{X}^-)^2$,
with the result
\be K_{X^{\ad \a} \, X_{\a \ad}
(X^-)^2 (\bar{X}^-)^2 }
= \left\{ -{\rm i}\, \Big(\frac{F}{{\rm e}^{s F} - {\bf 1}}
\Big)_{\a \ad}{}^{\ad \a}
 +  S_{\a \ad} \, S^{\ad \a} \right\} \,
K_{ (X^-)^2 (\bar{X}^-)^2 } ~;
\ee
see \cite{MG} for more details.

The linear differential
equation for $K_{(X^-)^2 (\bar{X}^-)^2}$ can be solved exactly,
and the solution will be given in a separate
publication.
Here we only reproduce the expansion
of $K^{--}$ to sixth order in spinor derivatives of $W$.
After making Schwinger's rotation $ s \rightarrow -{\rm i} \,\t$
of the proper-time parameter, $K^{--}$ reads
\bea
K^{--}(z;\t) &=&   \frac{\rm i}{32 \p^2} \, {\rm e}^{-  \t W \bar W} \,
(D^{+ } W \,D^+ W) \,
\Big\{ \bar W - \frac{{\rm i} \t}{2}(1 -
\frac{ \t}{2} \, W \bar W ) \, ({\bar D}_{\bd}^{ +} \bar W)
({\bar D}^{-\bd } \bar W) \non \\
&+ & \frac{\t^2}{24} \, \bar W \,
{\bar N}^{\ad}{}_{\bd} \, {\bar N}^{\bd}{}_{\ad}
+ \frac{\t^2}{6}(1 - \frac{ \t}{2} W \bar W) \, ({\bar D}_{\ad}^{ +}
\bar W )
\, {\bar N}^{\ad}{}_{\bd} \, ({\bar D}^{-\bd } \bar W) \non \\
& -& \frac{ \t^3}{16}\, W (1- \frac{\t}{4} W \bar W ) \,
({\bar D}^{ +} \bar W \, {\bar D}^{+} \bar W) \,
({\bar D}^{ -} \bar W \, {\bar D}^{- } \bar W ) \non \\
& + &\frac{ \t^3}{96}
({\bar D}^{ +} \bar W \, {\bar D}^{+ } \bar W ) \,
(D^{-} W \,D^{ -} W) \Big\}~ ~+~~ {\rm conjugate}~.
\eea
Then, performing the integral over $\t$ gives
\bea
 \d \G_{\rm H} = & - & \frac{1}{(8 \p)^2}
\int {\rm d} \z^{(-4)}\, \delta V^{++}
\, (D^{+} W \,D^+ W) \, \Big\{ \frac{1}{W}
+ \frac{1}{12} \,\frac{
\bar N^{\ad}{}_{\bd} \, \bar N^{\bd}{}_{\ad}
}{ W^3 \bar W^2} \,
\label{finalvar} \\
& - & \frac{1}{6} \,
\frac{
({\bar D}_{\ad}^{ +} \bar W )
\, \bar N^{\ad}{}_{\bd} \, ({\bar D}^{-\bd } \bar W )
}{(W \bar W )^3 }
+ \frac{1}{16} \,
\frac{
({\bar D}^{ +} \bar W \, {\bar D}^{+ } \bar W ) \,
({\bar D}^{ -} \bar W \, {\bar D}^{-} \bar W )
}{W^3 {\bar W}^4 }
\Big\}
~+~ {\rm c.c.} \non
\eea

Let us analyse the expression for $ \d \G_{\rm H}$ derived above.
The first term on the right hand side of (\ref{finalvar})
is quadratic in derivatives of $W$. This term, which
coincides with the earlier result (\ref{con2}),
is generated by the on-shell variation of
the holomorphic prepotential (\ref{holprep}).
The variation $ \d \G_{\rm H}$ does not involve
any terms of fourth order in derivatives of $W$ and $\bar W$,
although such terms do appear at intermediate
stages of the calculation. The reason all such terms
cancel out in the final expression for $ \d \G_{\rm H}$
is very simple. As was already mentioned, the effective
current $\cJ^{++}$ should be analytic. But there exists
no analytic superfield which  carries harmonic
$U(1)$ charge $+2$ and is fourth order in spinor
derivatives of $W$ and $\bar W$. Apart from the
first term on the right hand side of (\ref{finalvar}),
the rest of the terms are sixth order in derivatives of $W$ and $\bar W$.
They are  generated by the on-shell variation of
the leading non-holomorphic
correction in (\ref{structure}). Indeed,
under an arbitrary variation $V^{++} \to V^{++} + \d V^{++}$
of the analytic gauge field,
a real functional
$\int {\rm d}^{12}z\, \cH (W, \bar{W} )$ changes
as follows (see, e.g., \cite{DK}):
\bea
\delta  \int {\rm d}^{12}z\, \cH (W, \bar{W} )
= \frac{1}{64} \int {\rm d} \z^{(-4)}\, \delta V^{++} \,
(D^+)^2  (\bar D^+)^2 ({\bar D}^-)^2 \, \frac{\partial \cH (W, \bar
W )}{\partial W} ~+~ {\rm c.c.}
\label{nonhol}
\eea
 The terms in the  on-shell variation (\ref{finalvar})
of sixth order in spinor derivatives
 are generated by the superfield
\be
 \cH(W, \bar W) = \frac{1}{192 \p^2} \,
\ln \frac{W}{\L} \, \ln \frac{ \bar W}{\L}~.
\ee
This quantum correction was previously computed in
\cite{BKT} with the use of $\cN=1$ superfield techniques
and $\cN=2$ superconformal considerations.

\noindent
{\bf 5.} In this paper, we have presented a method for
computing the hypermultiplet effective action using
$\cN=2$ harmonic superspace heat kernel techniques.
Combined with the results of our earlier paper \cite{KM},
this provides a prescription for the analysis of
the effective action in generic $\cN=2$ super Yang-Mills
theories. Explicit calculations were presented only for
the case of $U(1)$ background; these can be extended
to arbitrary non-Abelian backgrounds in a fairly straightforward
manner. It would also
be of interest to develop a method to calculate
the hypermultiplet effective action directly rather
than by integrating its variation.

\vskip.5cm

\noindent
{\bf Acknowledgements:}
SMK is grateful to Professor Stefan Theisen for hospitality
at the Max Planck Institute for Gravitational Physics
(Albert Einstein Institute) where part of this work was done.
We thank Professor Joseph Buchbinder for useful comments.


\begin{thebibliography}{99}

\bi{WB} J. Wess and J. Bagger, {\it Supersymmetry and Supergravity},
Princeton Univ. Press, 1992 (2nd Edition).

\bi{GGRS} S.J.~Gates, M.T.~Grisaru, M.~Ro\v{c}ek and W.~Siegel,
{\it Superspace or One Thousand and One Lessons in Supersymmetry},
Benjamin-Cummings, Reading, MA, 1983.

\bibitem{BK} I.L. Buchbinder and S.M. Kuzenko,
{\it Ideas and Methods of Supersymmetry and
Supergravity or a Walk Through Superspace},
IOP, Bristol, 1998 (2nd Edition).

\bibitem{GIKOS} A. Galperin, E. Ivanov, S. Kalitzin,
V. Ogievetsky and E. Sokatchev,
Class. Quant. Grav. {\bf 1} (1984) 469.

\bibitem{GIOS}
A. Galperin, E. Ivanov, V. Ogievetsky and E. Sokatchev,
Class. Quant. Grav. {\bf 2} (1985) 601; 617.

\bibitem{BBKO} I.L. Buchbinder, E.I. Buchbinder,
S.M. Kuzenko and B.A. Ovrut,
Phys. Lett. {\bf B417} (1998) 61
[hep-th/9703147].

\bi{BK2} I.L.~Buchbinder and S.M.~Kuzenko,
Mod.\ Phys.\ Lett. {\bf A13} (1998) 1623 [hep-th/9804168];
see also I.~Buchbinder, S.~Kuzenko and B.~Ovrut,
{\it Covariant harmonic supergraphity for $N = 2$ super Yang-Mills theories},
in {\it Supersymmetries and Quantum Symmetries},
J. Wess and E.A. Ivanov eds., Springer, 1999, p. 21,
hep-th/9810040.

\bibitem{BKO}
I.L.~Buchbinder, S.M.~Kuzenko and B.A.~Ovrut,
Phys.\ Lett.\ {\bf B433} (1998) 335
[hep-th/9710142].

\bibitem{BBK}
E.I. Buchbinder, I.L. Buchbinder and S.M. Kuzenko,
Phys.\ Lett.\ {\bf B446} (1999) 216
[hep-th/9810239].

\bibitem{KM}
S.M.~Kuzenko and I.N.~McArthur,
Phys.\ Lett.\  {\bf B506} (2001) 140
[hep-th/0101127].

\bi{GSW} R. Grimm, M. Sohnius and J. Wess,
Nucl. Phys. {\bf B133} (1978) 275.

\bi{Z} B. Zupnik, Phys.\ Lett.\ {\bf B183} (1987) 175.

\bibitem{BBIKO}
I.L.~Buchbinder, E.I.~Buchbinder, E.A.~Ivanov, S.M.~Kuzenko
and B.A.~Ovrut,
Phys.\ Lett.\  {\bf B412} (1997) 309
[hep-th/9703147];
E.I.~Buchbinder, I.L.~Buchbinder, E.A.~Ivanov and S.M.~Kuzenko,
Mod.\ Phys.\ Lett.\  {\bf A13} (1998) 1071
[hep-th/9803176].

\bibitem{DK}
N.~Dragon and S.M.~Kuzenko,
Nucl.\ Phys.\  {\bf B508} (1997) 229
[hep-th/9705027].

\bibitem{IKZ}
E.A.~Ivanov, S.V.~Ketov and B.M.~Zupnik,
Nucl.\ Phys.\ B {\bf 509} (1998) 53
[hep-th/9706078].

\bibitem{GR}
F.~Gonzalez-Rey and M.~Ro\v{c}ek,
Phys.\ Lett.\ {\bf B434} (1998) 303,
hep-th/9804010.

\bibitem{BKT}
I.L.~Buchbinder, S.M.~Kuzenko and A.A.~Tseytlin,
Phys.\ Rev.\  {\bf D62} (2000) 045001
[hep-th/9911221].

\bibitem{SW}
N.~Seiberg and E.~Witten,
Nucl.\ Phys.\ {\bf B426} (1994) 19
[Erratum-ibid.\ {\bf B430} (1994) 485]
[hep-th/9407087];
Nucl.\ Phys.\ {\bf B431} (1994) 484
[hep-th/9408099].

\bibitem{DS}
M.~Dine and N.~Seiberg,
Phys.\ Lett.\ {\bf B409} (1997) 239
[hep-th/9705057].

\bi{superconformal} A. Galperin, E. Ivanov, V. Ogievetsky
and E. Sokatchev, {\it Conformal invariance in harmonic superspace},
in {\it Quantum Field Theory and Quantum Statistics},
I.A. Batalin, G.A. Vilkovisky and C.J. Isham eds., Hilger, Bristol,
1987, p. 233;
P.S.~Howe and G.G.~Hartwell,
Class.\ Quant.\ Grav.\  {\bf 12} (1995) 1823;
S.M.~Kuzenko and S.~Theisen,
Class.\ Quant.\ Grav.\  {\bf 17} (2000) 665
[hep-th/9907107].

\bibitem{hol}
S.~Eremin and E.~Ivanov,
Mod.\ Phys.\ Lett.\  {\bf A15} (2000) 1859
[hep-th/9908054];
I.L.~Buchbinder and I.B.~Samsonov,
Mod.\ Phys.\ Lett.\  {\bf A14} (1999) 2537
[hep-th/9909183].

\bibitem{BKY}
I.L.~Buchbinder, S.~Kuzenko and Z.~Yarevskaya,
Nucl.\ Phys.\  {\bf B411} (1994) 665.

\bibitem{DGR}
B.~de Wit, M.T.~Grisaru and M.~Ro\v{c}ek,
Phys.\ Lett.\  {\bf B374} (1996) 297
[hep-th/9601115];
M.T.~Grisaru, M.~Ro\v{c}ek and R.~von Unge,
Phys.\ Lett.\  {\bf B383} (1996) 415
[hep-th/9605149].

\bibitem{PW}
A.~Pickering and P.~West,
Phys.\ Lett.\  {\bf B383} (1996) 54
[hep-th/9604147].

\bibitem{PB}
N.G.~Pletnev and A.T.~Banin,
Phys.\ Rev.\  {\bf D60} (1999) 105017
[hep-th/9811031];
A.T.~Banin, I.L.~Buchbinder and N.G.~Pletnev,
Nucl.\ Phys.\ {\bf B598} (2001) 371
[hep-th/0008167].


\bibitem{MG}
I.N.~McArthur and T.D.~Gargett,
Nucl.\ Phys.\  {\bf B497} (1997) 525
[hep-th/9705200];
T.D.~Gargett and I.N.~McArthur,
J.\ Math.\ Phys.\  {\bf 39} (1998) 4430.



\end{thebibliography}
\end{document}